\documentclass[epj]{webofc}
\usepackage[varg]{txfonts}   
\wocname{EPJ Web of Conferences}

\woctitle{INPC 2013}

\begin{document}

\title{Up- and Down-Quark Contributions to the Nucleon Form Factors}

\author{I. A. Qattan\inst{1}\fnsep\thanks{\email{issam.qattan@kustar.ac.ae}} \and
        J. Arrington\inst{2}
}

\institute{Department of Applied Mathematics and Sciences, Khalifa University
of Science, Technology and Research, P.~O. Box 573, Sharjah, U.A.E \and Physics
Division, Argonne National Laboratory, Argonne, Illinois 60439, USA
}

\abstract{
Recent measurements of the neutron's electric to magnetic form factors ratio, $R_{n}= \mu_{n} G_E^n/G_M^n$, up to 3.4 (GeV/c)$^2$ combined with existing $R_{p}= \mu_{p} G_E^p/G_M^p$ measurements in the same $Q^2$ range allowed, for the first time, a separation of the up- and down-quark contributions to the form factors at high $Q^2$, as presented by Cates, {\it{et al.}}. Our analysis expands on the original work by including additional form factor data, applying two-photon exchange (TPE) corrections, and accounting for the uncertainties associated with all of the form factor measurements.}

\maketitle


The proton's elastic form factors, $G_E^p$ and $G_M^p$, provide information on the spatial distributions of charge and magnetization of the nucleon. In the nonrelativistic limit, they are simply the Fourier transform of the charge and magnetization distributions.  Thus, isolating the up- and down-quark contributions can be used to examine spatial asymmetries in the quark distributions, just as the differences between $G_E^p$ and $G_M^p$~\cite{puckett2012} indicate a difference between the charge and magnetization distributions~\cite{kelly2002}.

The nucleon form factors can be extracted in unpolarized elastic scattering, using the longitudinal-transverse (LT) or Rosenbluth separation technique, or in polarization transfer/polarized target (PT) measurements~\cite{rosen_dombey}.  The unpolarized cross section in the one-photon exchange (OPE) is proportional to the so-called `reduced' cross section, $\sigma_{R}= G_M^2+(\varepsilon/\tau) G_E^{2}$, where $\tau=Q^2/4M_{N}^2$, $M_N$ is the nucleon mass, and $\varepsilon$ is the virtual photon longitudinal polarization parameter.  By varying $\varepsilon$ (related to the scattering angle) at a fixed $Q^2$, one can separate $G_E$ and $G_M$.  For cases where $\varepsilon/\tau$ is extremely small (large), it is difficult to extract $G_E$ ($G_M$) with precision. The polarization measurements are sensitive only to the ratio $G_E/G_M$, and thus complement the cross section measurements in regions where the cross section is dominated by one of the form factors.  By taking ratios of polarization components, many of the systematic uncertainties in the polarization measurements cancel, allowing for precise measurements of $R_p = \mu_{p} G_E^p/G_M^p$~\cite{puckett2012}.  The two methods yield significantly and strikingly different results for the $G_E^p/G_M^p$ in the region $Q^2 > 2$~(GeV/c)$^2$~\cite{arrington03}. The Rosenbluth extractions show approximate scaling, $\mu_{p} G_E^p/G_M^p \approx 1$, while the recoil polarization data indicate a nearly linear decrease in $R_p$ with $Q^2$. Recent studies suggest that hard two-photon exchange (TPE) corrections to the unpolarized cross section may resolve the discrepancy~\cite{arrington07, CV2007, ABM2011}.

Separation of the up- and down-quark contributions at large $Q^2$ was recently made possible by precise measurements of the neutron form factor ratio $R_{n} = \mu_{n} G_E^n/G_M^n$ up to 3.4~GeV$^2$~\cite{riordan2010}.  The first analysis of the flavor-separated form factors by Cates \textit{et al.}~\cite{CJRW2011}, referred to as ``CJRW'' in this work, emphasized the differences in the scaling behavior of the up- and down-quark contributions at large $Q^2$, supporting the idea that diquark correlations play a significant role~\cite{cloet09}.  In our recent work~\cite{qattan2012}, we extended the flavor separation analysis, accounting for effects neglected in the original work.  We include the impact of TPE corrections in the extraction of the proton form factors, following the approach of Ref.~\cite{qattan2011}. This takes the TPE parametrization from Ref.~\cite{borisyuk-kobushkin2011}, which assumes that the corrections are linear in $\varepsilon$~\cite{tvaskis2006} and vanish in the limit of small angle scattering~\cite{arrington11-chen07,arrington04}. Because this approach is less reliable at low $Q^2$, we compare to results using a parametrization of the proton form factors extracted after applying the hadronic calculation for TPE~\cite{arrington07}.  In addition, we include estimates for the uncertainties for all of the form factors while the CJRW analysis included only the uncertainties on $R_n$ (which yield the dominant uncertainty for many of the flavor-separated results).  Finally, we include additional form factor data, in particular the new $G_M^n$ data from CLAS~\cite{lachniet09}.  We note that there is significant tension between different extractions of $G_M^n$ for $Q^2 \approx 1$~GeV$^2$, and the uncertainties taken do not account for the discrepancy in different data sets.~\cite{qattan2012}.

We compare our results to the CJRW analysis, to examine the impact of the TPE corrections, additional uncertainties, and updated form factor data set, and to recent proton form factor parametrization of Venkat {\textit{et al.}}~\cite{venkat2011} ("VAMZ"), and Arrington {\textit{et al.}}~\cite{arrington07} ("AMT"), to examine the impact of the different TPE prescriptions at lower $Q^2$ values. We also study the effect of the updated fit to $G_M^n$ by showing a version of the VAMZ extraction which uses the Kelly~\cite{kelly2004} parametrization for $G_M^n$ ("VAMZ-Kelly") rather than our updated $G_M^n$ parametrization, which shows a noticeable difference near $Q^2=1$~GeV$^2$. Finally, we show recent calculations and fits to the flavor-separated contributions: a Dyson-Schwinger equation ("DSE") calculation~\cite{cloet09}, a pion-cloud relativistic constituent quark model ("PC-RCQM")~\cite{cloet2012}, a relativistic constituent quark model whose hyperfine interaction is derived from Goldstone-boson exchange ("GBE-RCQM")~\cite{rohrmoser}, and a generalized parton distribution (GPD) calculations~\cite{gonzalez}.  Note that the GPD and PC-RCQM models involve fitting a significant number of parameters to both the form factor measurements and other observables, and so have significant flexibility to fit the data, while the DSE and PC-RCQM results do not adjust any parameters to match the form factor data.

\begin{figure}
\includegraphics[width=4.66cm,clip]{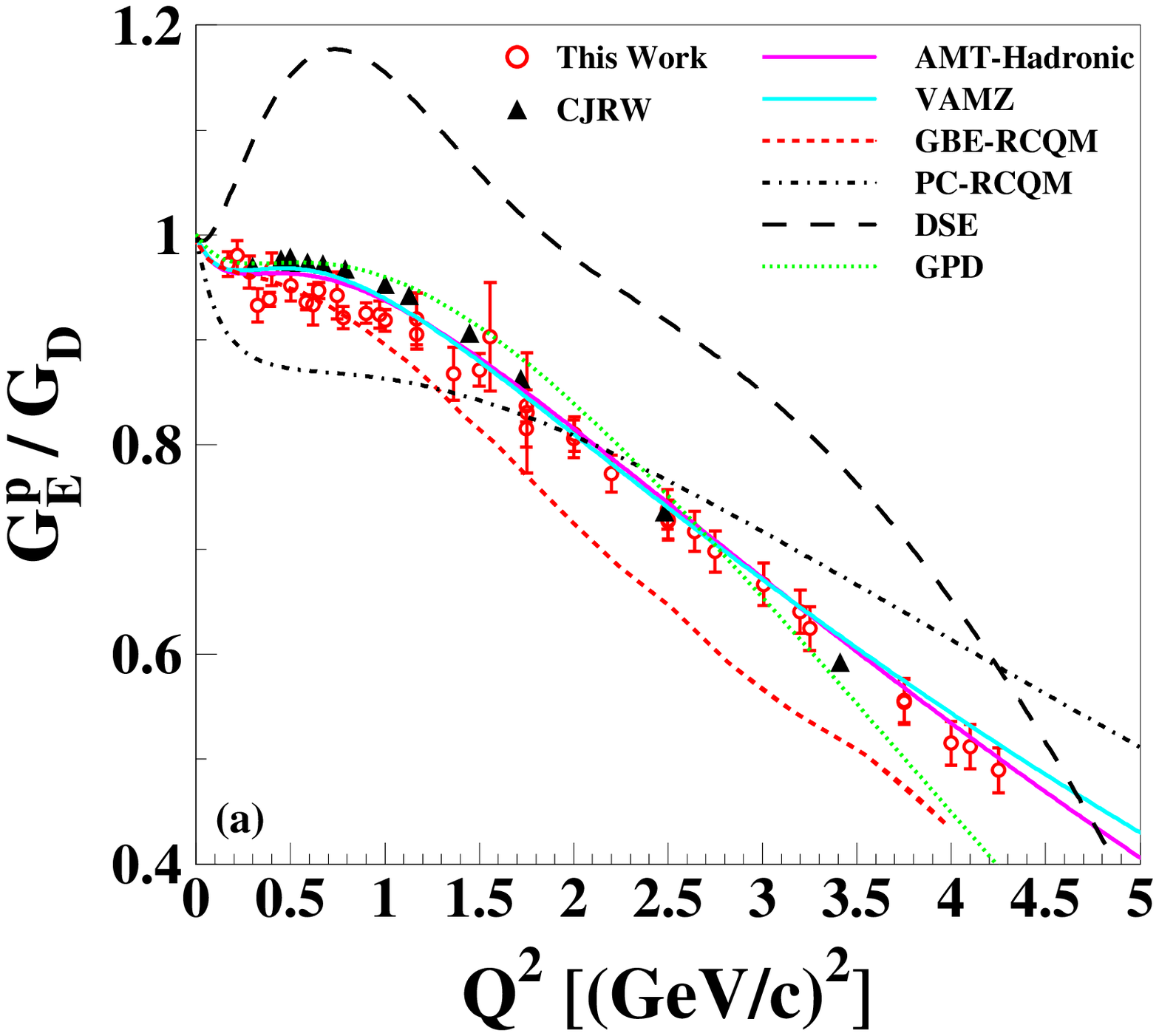}
\includegraphics[width=4.66cm,clip]{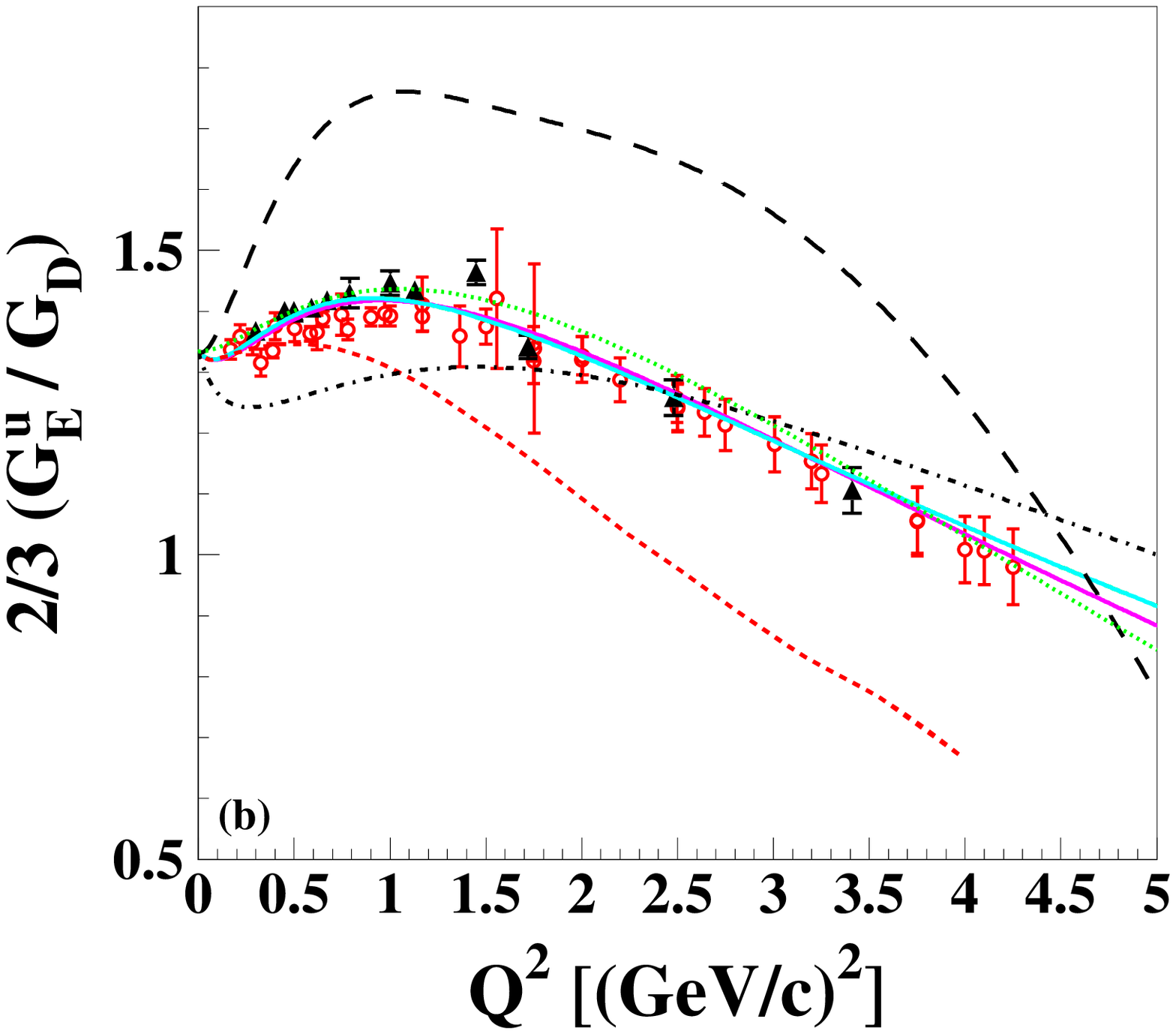}
\includegraphics[width=4.66cm,clip]{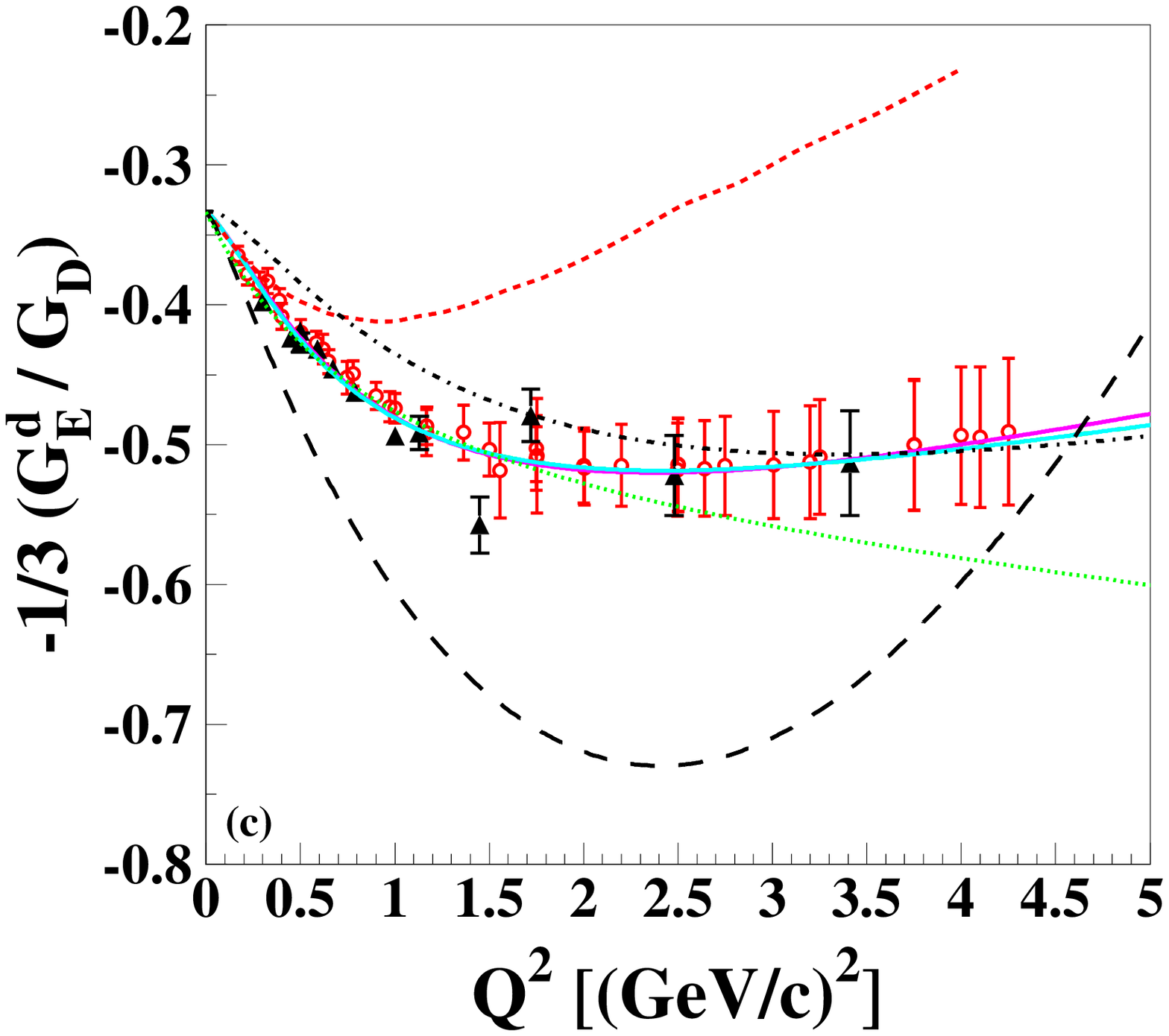}\\
\includegraphics[width=4.66cm,clip]{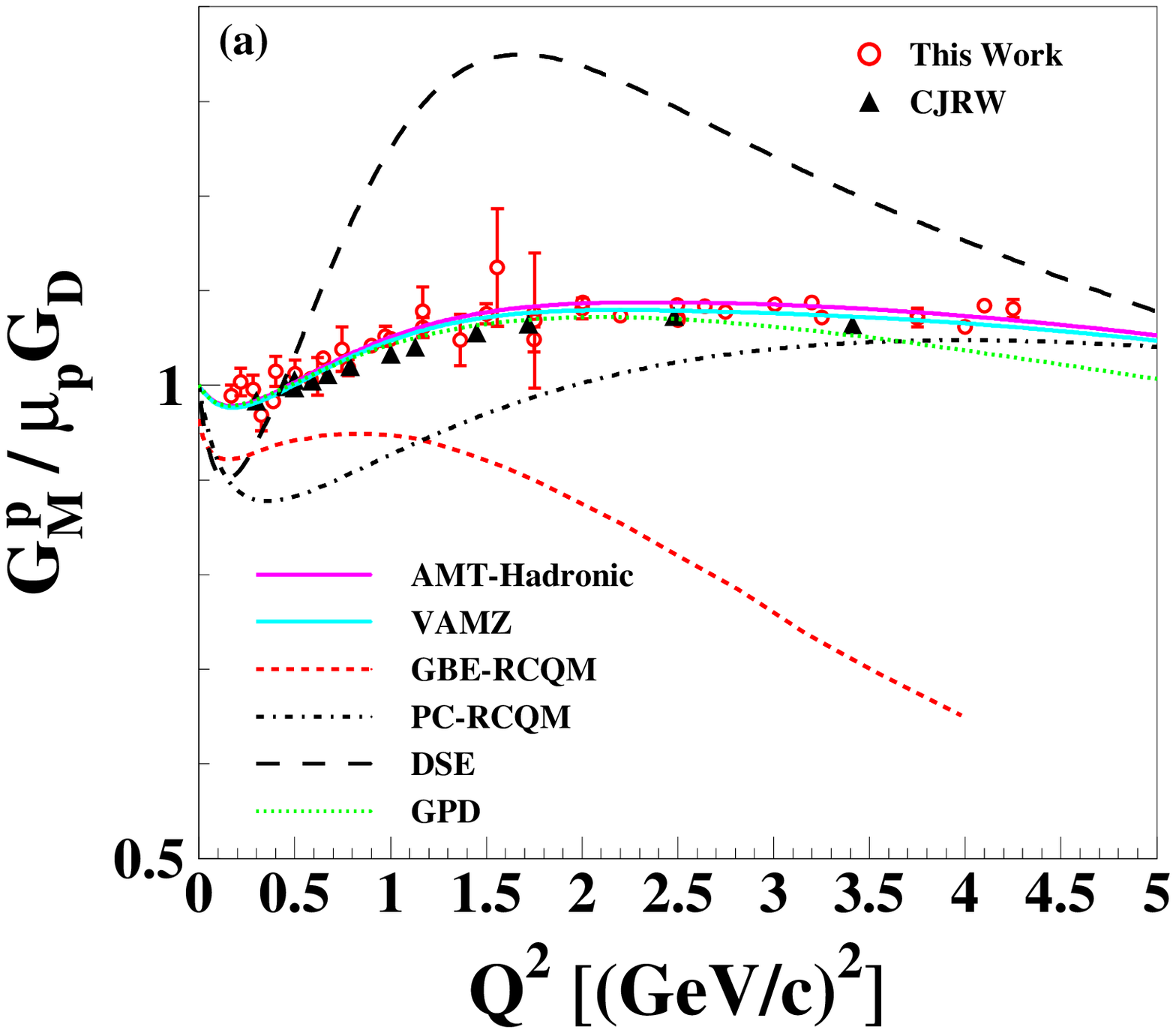}
\includegraphics[width=4.66cm,clip]{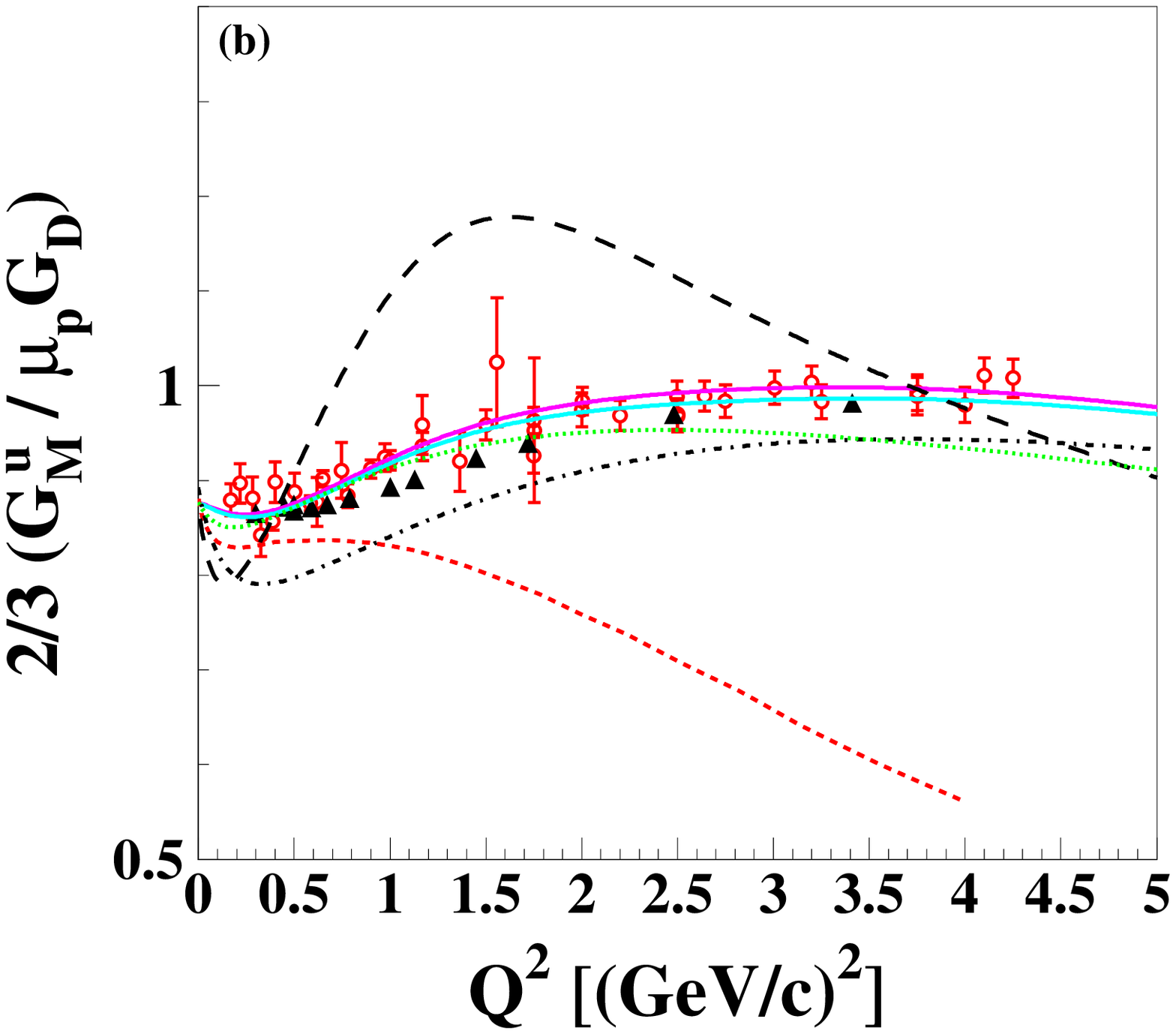}
\includegraphics[width=4.66cm,clip]{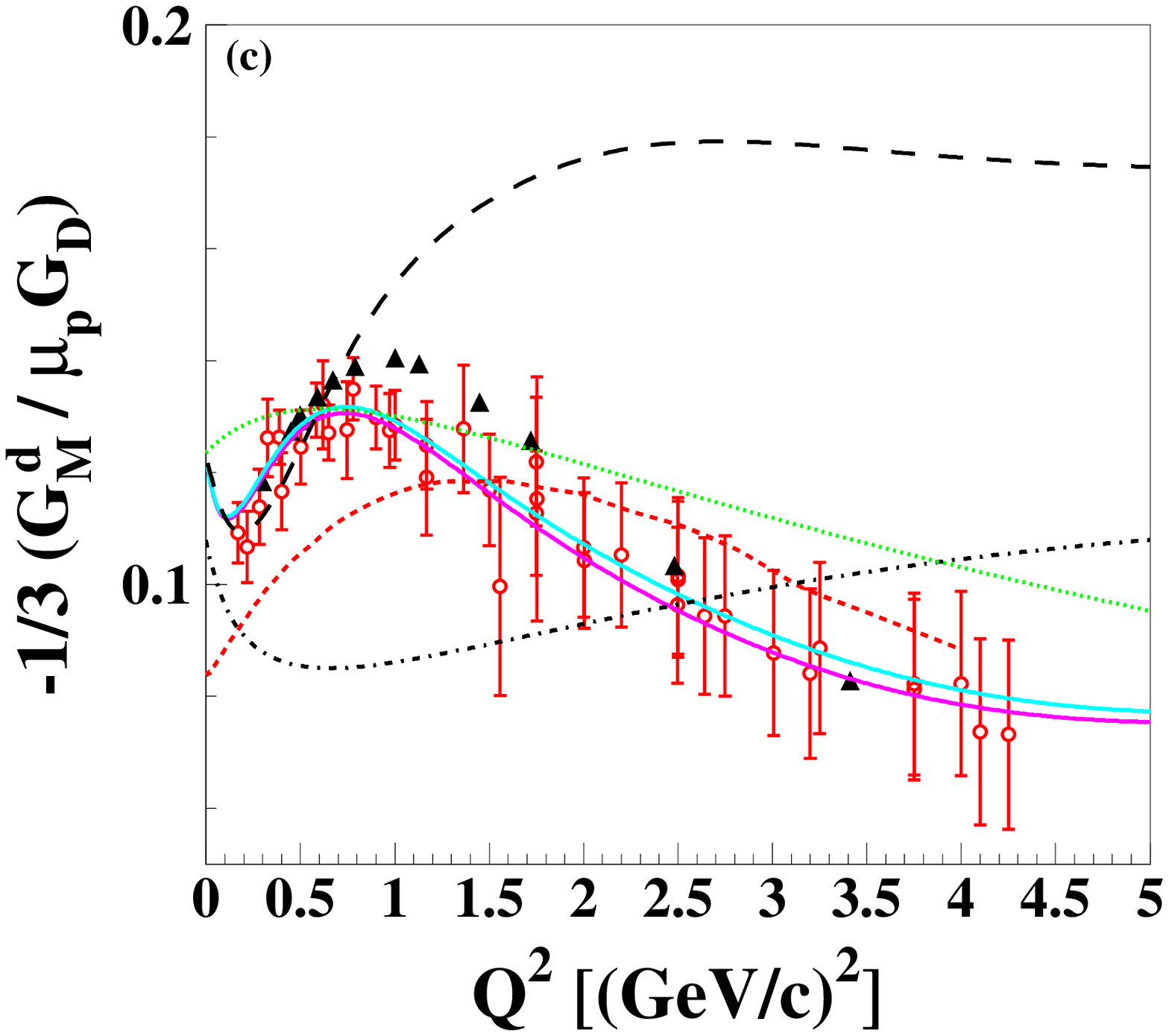}
\caption{$G_E^p$ (top) and $G_M^p$ (bottom) normalized to the dipole form, along with their up- and down-quark contributions from our analysis~\cite{qattan2012} and the CJRW extractions~\cite{CJRW2011}. Also shown are the AMT~\cite{arrington07} and VAMZ fits~\cite{venkat2011}, and the values from the GBE-RCQM~\cite{rohrmoser}, PC-RCQM~\cite{cloet2012}, the DSE~\cite{cloet09}, and the GPD~\cite{gonzalez} models.}
\label{fig1}
\end{figure}

\begin{figure}
\includegraphics[width=4.66cm,clip]{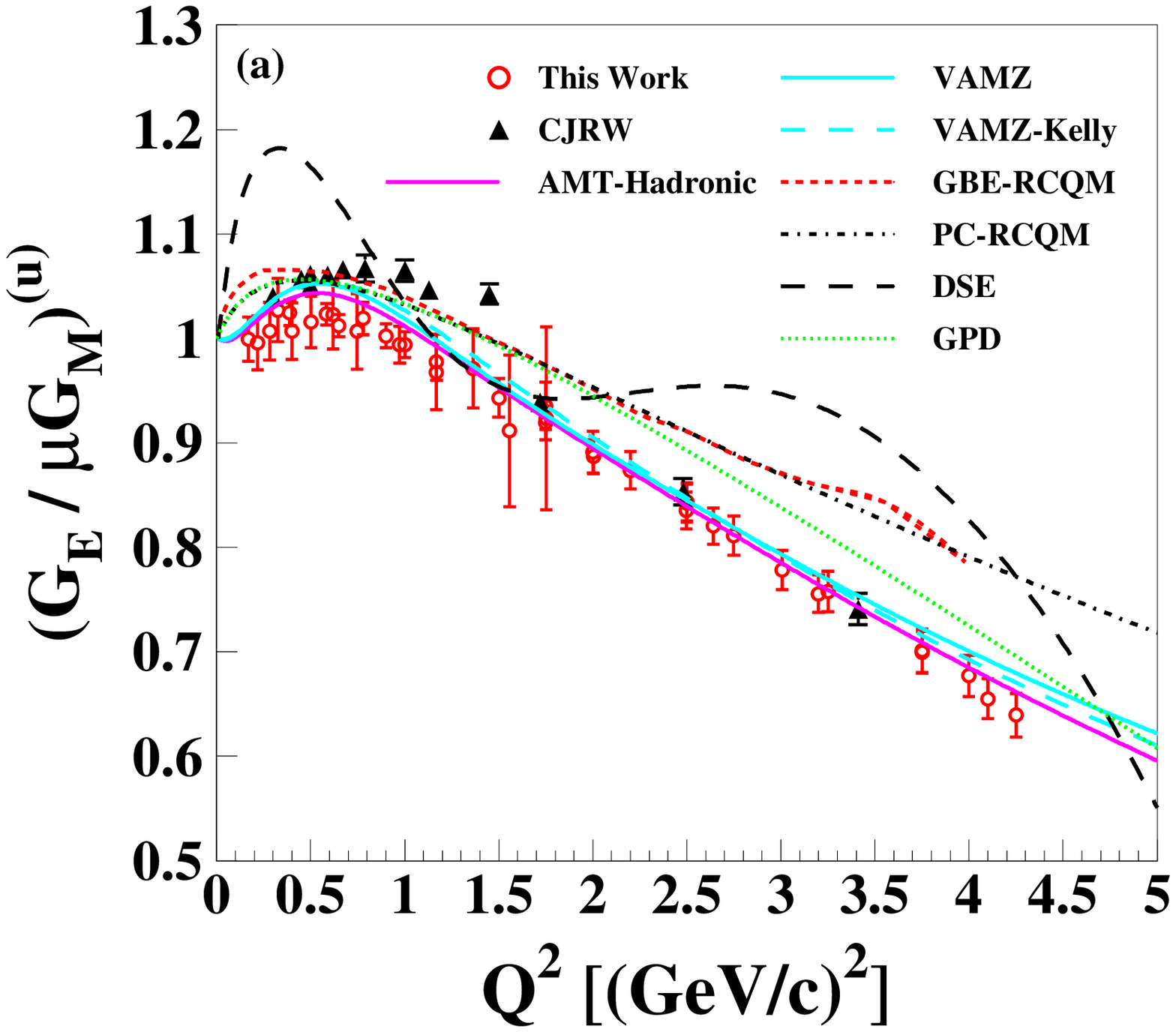}
\includegraphics[width=4.66cm,clip]{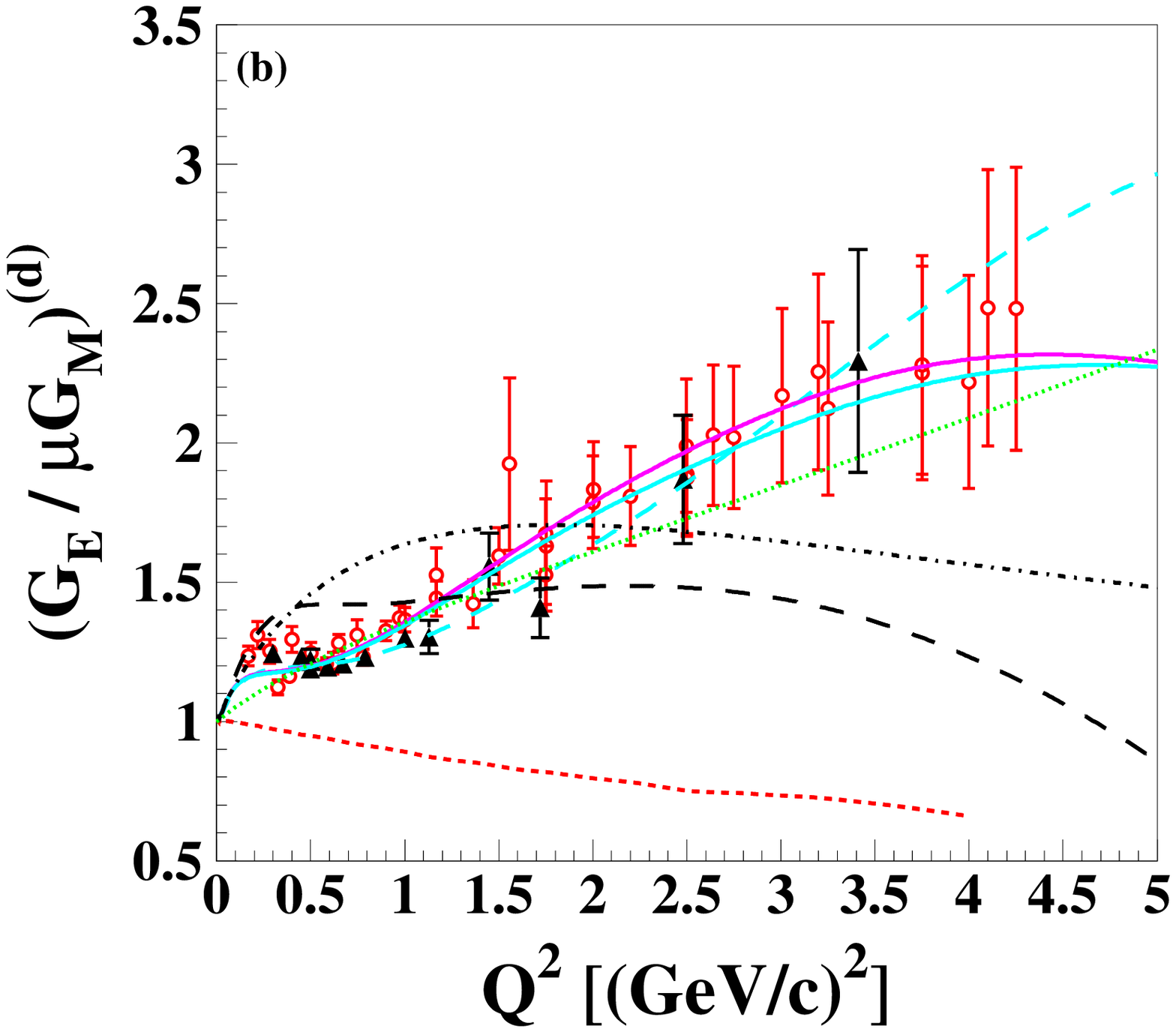}
\includegraphics[width=4.66cm,clip]{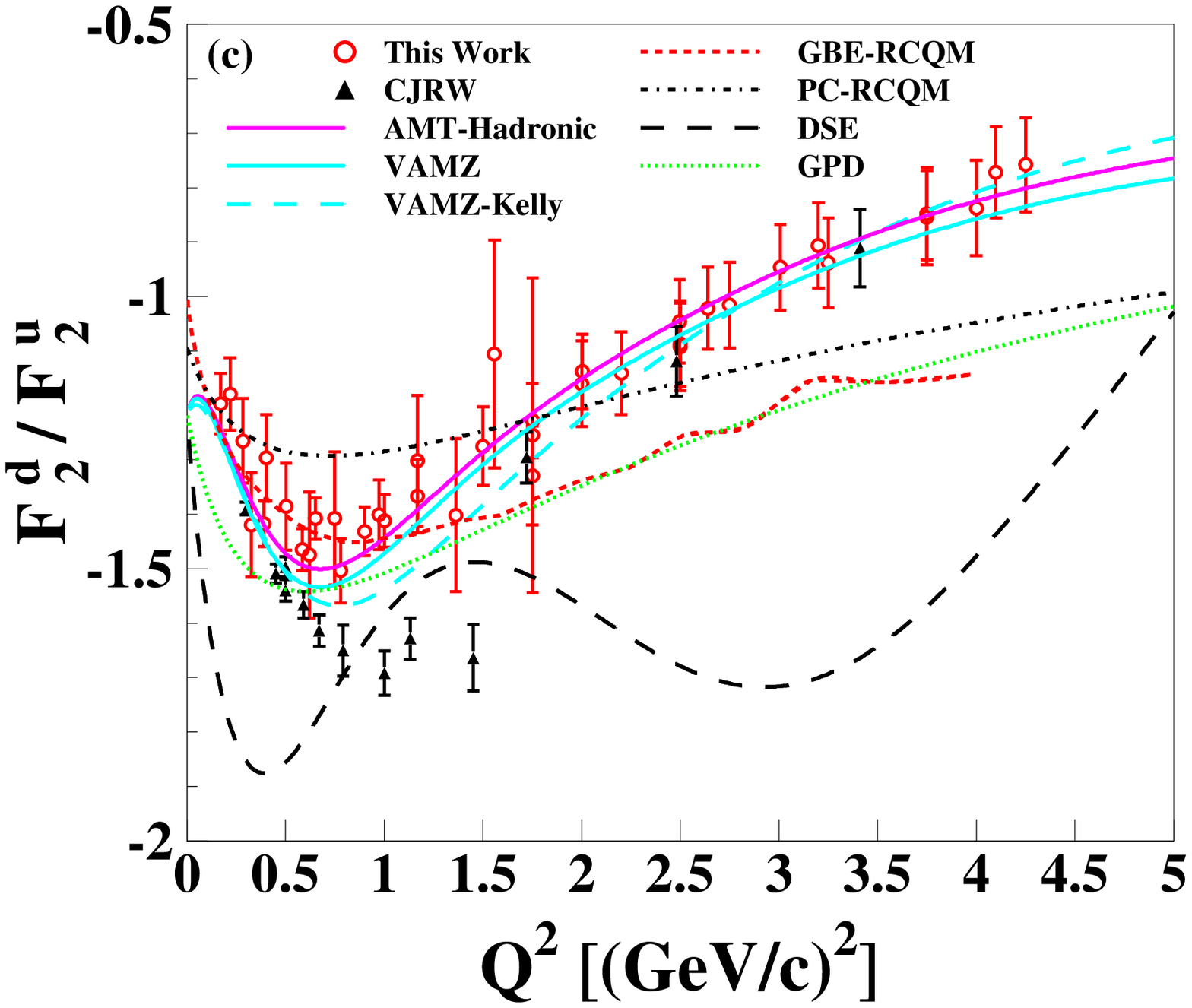}
\caption{Flavor separated contributions to $G_E/\mu G_M$, and the ratio $F_2^d/F_2^u$.  Points and curves as in Fig.~\ref{fig1}.}
\label{fig2}
\end{figure}


Figure~\ref{fig1} shows the Sachs form factors of the proton, along with their up- and down-quark contributions. Except for the overall normalization factors, the up-quark contribution to the proton is identical to the down-quark contribution to the neutron and vice-versa.  Note that the CJRW results for the magnetic form factor contributions and $G_E^p$ have no uncertainty as only the uncertainty on $R_n$ was included, and $R_n$ does not enter into these observables. As expected, the up-quark contribution dominates for the proton. In addition, there is a clear difference between our results and the CJRW extraction in some observables for $Q^2 \approx 1$~GeV$^2$, caused by a combination of the TPE corrections to the proton and the modified fit to the neutron magnetic form factor. The up-quark contribution to $G_E$ falls slowly, while the down quark contribution grows or is approximately constant over the $Q^2$ range.  Because of the negative contribution of the down quark, it is the combination of these behaviors that yields the rapid falloff of the proton charge form factors.  For the magnetic form factor, both the up and down quark contributions have relatively small deviations from the dipole form, but these partially cancel in the sum, yielding a more constant value for $G_M^p/G_D$.  Thus, the behavior of $G_M^p$ is in almost perfect agreement with the dipole form, despite the fact that both the up and down quarks show significant deviations.  The differences between the extracted points and the extractions based on calculated TPE corrections (VAMZ and AMT-hadronic), seen most clearly in $G_E^p$, suggest that there is a small difference between the phenomenological TPE extraction used here and the calculated correction at low $Q^2$, as expected~\cite{qattan2012}.  The GPD and PC-RCQM do a better job describing the qualitative results, but remember that these include the proton and neutron form factors in their fits.  So while the DSE and GBE-RCQM results do not reproduce the data at the same level, they are true predictions for the flavor-separated contributions and their agreement is rather impressive.

Figure~\ref{fig2} shows the ratio $G_E/\mu G_M$ for both the up- and down-quark contributions, along with the ratio of the down- and up-quark contributions to the Pauli form factor, $F_2$. In the $G_E/G_M$ ratios we can again see that the up- and down-quark contributions not only differ from each other, but are also significantly different from the nearly linear falloff observed in $G_E^p/G_M^p$~\cite{puckett2012}.  It is also clear that the calculations do not reproduce the detailed behavior.  The comparison of the VAMZ and VAMZ-Kelly parameterizations show the impact of the modified parametrization of $G_M^n$, which has a noticeable impact on the down quark ratio at higher $Q^2$ values, due to the difference between the CLAS high-$Q^2$ behavior and the previous fit which had little data to constrain the high-$Q^2$ behavior.  In the ratio $F_2^d/F_2^u$, there is a small difference near $Q^2=1$ due to the modified $G_M^n$ parametrization, but an even larger difference between this work and the CJRW extraction, which is a combination of the modified neutron form factor and the TPE corrections.  Note that in this ratio, none of the calculations yield a good description of the data, making this a particularly useful place to examine the quality of the models of the nucleon form factors and their flavor-dependent contributions.


The flavor-separated form factors clearly have significant power to test models of the nucleon structure, and will provide important input to models of generalized parton distributions.  Additional measurements of $G_E^n$ at higher $Q^2$ values will be performed after the JLab 12 GeV upgrade is completed, providing further information on the high-$Q^2$ behavior of the form factors and the difference in the up- and down-quark contributions.  In addition, these results suggest that further measurements to clarify the neutron magnetic form factor at lower $Q^2$ would also improve the extraction of the down-quark contributions.

This work was supported by Khalifa University of Science, Technology and Research and by the U.S. Department of Energy, Office of Nuclear Physics, under contract DE-AC02-06CH11357.

%
%

\end{document}